\documentclass[final,3p,times,12pt,authoryear]{elsarticle}

\usepackage{graphicx}
\usepackage{amssymb}

\usepackage{lineno}

\usepackage{gensymb}

\usepackage{amsmath}

\usepackage{bm}

\usepackage[labelformat=empty]{subfig}

\usepackage{color,soul}

\usepackage{textcomp}

\journal{Journal of the Mechanics and Physics of Solids}

\begin{document}

\begin{frontmatter}


\title{Characterisation of slip and twin activity using digital image correlation and crystal plasticity finite element simulation: Application to orthorhombic $\alpha$-uranium}



\author[engineering]{Nicol\`{o} Grilli}
\author[materials]{Philip Earp}
\author[engineering]{Alan C.F. Cocks}
\author[materials]{James Marrow}
\author[materials]{Edmund Tarleton \corref{cor}}

\ead{edmund.tarleton@materials.ox.ac.uk}

\cortext[cor]{Corresponding author}
\address[engineering]{Department of Engineering Science, University of Oxford, Parks Road, OX1 3PJ, UK}
\address[materials]{Department of Materials, University of Oxford, Parks Road, OX1 3PH, UK}

\begin{abstract}

Calibrating and verifying crystal plasticity material models is {a} significant challenge, particularly for materials with a number of potential slip and twin systems. Here we use digital image correlation on coarse-grained $\alpha$-uranium  during tensile testing in conjunction with crystal plasticity finite element simulations. This approach allows us to determine the critical resolved shear stress, and hardening rate of the different slip and twin systems. The constitutive model is based on dislocation densities as state variables and the simulated geometry is constructed from electron backscatter diffraction images that provide shape, size and orientation of the grains, allowing a direct comparison between virtual and real experiments. An optimisation algorithm is used to find the model parameters that reproduce the evolution of the average strain in each grain as the load is increased. A tensile bar, containing four grains aligned with the load direction, is used to calibrate the model with eight unknown parameters. The approach is then independently validated by simulating the strain distribution in a second tensile bar. Different mechanisms for the hardening of the twin systems are evaluated. The latent hardening of the most active twin system turns out to be determined by coplanar twins and slip. The hardening rate of the most active slip system is lower than in fine-grained $\alpha$-uranium. The method developed in the present research can be applied to identify the critical resolved shear stress and hardening parameters of other coarse-grained materials. 
\end{abstract}

\begin{keyword}
Crystal Plasticity \sep Dislocations \sep Twinning \sep Tensile test \sep EBSD \sep Digital Image Correlation \sep Uranium


\end{keyword}

\end{frontmatter}


\section{Introduction}
\label{S:Introduction}

Hardening in metals is due to interactions within, and between, the slip and twin systems \citep{KALIDINDI1998267}. The nucleation and motion of dislocations and twin nucleation and migration {occur} when the critical resolved shear stress (CRSS) for these deformation mechanisms is reached \citep{Devincre1745,Ojha2014}. {Since the seminal research of} \citep{Asaro1983}, {there has been significant effort dedicated towards the development of crystal based models for the inelastic deformation of engineering materials.} {Models of this type can contain a large number of material parameters, particularly if there are a number of different slip and twin systems that can be activated.} {A challenge then is to develop robust experimental procedures for the determination of these parameters.} {This either requires conducting a series of different experiments on polycrystalline materials covering a range of loading conditions and or stress states, or testing single crystals with different crystallographic configurations with respect to the loading direction} \citep{FRANCIOSI197923}. {For a wide range of materials, particularly if they have limited ductility, each of these approaches might be impractical or even impossible to undertake.} {This is the case for the primary material of interest here - $\alpha$-uranium} \citep{ORNLreportUandhydrogen,Lander1994review}.  

{In this paper we demonstrate how the material parameters can be determined from a single uniaxial test on a coarse grained material by comparing the detailed strain and displacement fields within the individual grains determined using digital image correlation (DIC) with crystal plasticity finite element (CPFE) simulations.}  {For $\alpha$-uranium, this allows us to determine the 8 material parameters that describe the response of the different slip and twin systems that are active at room temperature.} {The approach can be readily generalised to other material systems.} 

Polycrystalline samples with columnar grains and arbitrary boundary conditions can be modelled using the crystal plasticity finite element (CPFE) method \citep{ROTERS2018}, which considers the anisotropic elastic behaviour, plastic deformation due to slip and twinning. Given the grain orientation, CPFE simulations provide the strain distribution in each grain \citep{DUNNE20071061} and can be compared to digital image correlation (DIC) measurements \citep{LIM20141}.

The CPFE method has been widely used in conjunction with DIC measurements. For instance, DIC measurements can determine the local strain in polycrystalline samples, which is used for small scale CPFE simulations \citep{NELLESSEN201586,GRILLI2018104}. CPFE simulations are able to predict the strain localization \citep{IRASTORZALANDA2016184} and lattice rotation \citep{IRASTORZALANDA2017157} in single crystals and oligocrystals, as detected by high resolution DIC \citep{GUAN201770}. The DIC results can be correlated with the presence of geometrically necessary dislocations near grain boundaries \citep{Roters2004}. DIC has also been used to correlate strain concentrations with fracture paths \citep{KhanMarrow2009}. As shown in the present study, the comparison between CPFE simulations and DIC measurements allows us to find the CRSS and hardening coefficients for the plastic deformation mechanisms in coarse-grained $\alpha$-uranium.

In the CPFE method, the CRSS and hardening of the slip systems can be described by dislocation-based models, where the dislocation densities on the slip systems are the state variables \citep{GRILLI2015424}. An increase in dislocation density leads to hardening, usually described using a matrix detailing the strength of interactions between slip systems \citep{ARSENLIS20021979}. The dislocation density increase in $\alpha$-uranium during deformation has not been systematically measured and few discrete dislocation dynamics studies appear in the literature \citep{Behmer2018ThesisDDD}, which focus on the effect of twin thickness on dislocation pile-up and yield stress. The CRSS for slip and twinning have been identified for single crystal \citep{Daniel1971SingleCrystal} and fine-grained $\alpha$-uranium \citep{McCabeTome2010}, obtained by hot or cold rolling. The interaction matrix between slip systems has been calibrated for fine-grained $\alpha$-uranium \citep{McCabeTome2010} using polycrystal simulations and the visco-plastic self-consistent (VPSC) homogenization scheme \citep{Hutchinson1970EPSC,LEBENSOHN19932611}. However, the CRSS and hardening have not been widely studied for coarse-grained $\alpha$-uranium.

In the CPFE method, the interaction between twins and dislocations can be described using a CRSS for twin systems that depends on the accumulated shear on the slip systems \citep{AbdolvandDaymond2JMPS2013}. The hardening rate of existing twins is also governed by the activity of coplanar and non-coplanar twins \citep{ROTERS20101152}. This is based on the observation that, in FCC metals, deformation twinning initially produces only coplanar twins, while non-coplanar twin variants appear at larger strains \citep{RKALIDINDI2001837}. This has led to the introduction of self hardening and latent hardening coefficients that couple the CRSS for twinning with the twin volume fraction \citep{AbdolvandDaymond2JMPS2013}. These coefficients are set to zero in constitutive models describing fine-grained $\alpha$-uranium, calibrated using stress-strain curves \citep{McCabeTome2010} and twin volume fraction measurements \citep{KNEZEVIC2012702}. However, polycrystal simulations and fine-grained samples, in which thousands of grains are considered and the twin volume fraction is low \citep{ORNLreportUandhydrogen}, are not ideal to study twin-slip and twin-twin interactions. The same holds true for neutron diffraction experiments \citep{Earp2018Texture,Grilli2019Uranium,GrilliCOMPLAS2019}, in which the measured lattice strain is averaged over a millimetre-sized gauge volume that contains many grains.

Coarse-grained $\alpha$-uranium, obtained by casting, has a base centred orthorhombic crystal structure \citep{JacobWarren1937,Lukesh1949,Kapoor2015HotDeformationABG} and grains reaching a size of several millimetres \citep{Garlea2013HighStrainRate}. It is ideal to study slip-twin interactions because of the propensity to form several twins after quenching or deformation \citep{Cahn1951,CAHN195349}. This is because the CRSS for twinning is comparable to the CRSS {for} slip. Plastic deformation mechanisms in $\alpha$-uranium are strongly influenced by temperature \citep{Daniel1971SingleCrystal,Lloyd1966CTE}, strain rate \citep{HUDDART1980316}, grain size \citep{Taplin1967,ORNLreportUandhydrogen}, grain shape \citep{Taplin1963GrainShape}, loading conditions \citep{Zecevic2016Texture}, hydrogen content \citep{Calhoun2015Hydrogen} and alloying elements \citep{YANZHI20151094,Chen2017IndentationUNb}. At room temperature, the most active slip system is the so-called ``wall'' slip system, which has the lowest CRSS up to about 400$\degree$C, followed by the ``floor'' slip system \citep{ZECEVIC2016143}. Other slip systems exist, such as the ``chimney'' slip system, with two variants, and the ``roof'' slip system, with four variants \citep{Calhoun2018T150C,Calhoun2016Thesis,CalhounAgnew2013ThermalStrains}, but are observed only at high temperatures ($T>873$ K) when the temperature is close to the $\alpha \rightarrow \beta$ phase transition \citep{Daniel1971SingleCrystal,Kapoor2015HotDeformationABG}. The most commonly observed twin system is $\{130\}\langle3\bar{1}0\rangle$ \citep{Brown2009NeutronDiffraction}, but a small fraction of \textquotesingle$\{172\}$\textquotesingle$\langle3\bar{1}2\rangle$ twins are also observed \citep{Zhou2016retwinning}. The \textquotesingle$\{176\}$\textquotesingle$\langle512\rangle$ twin system is observed only at high strain rate \citep{HoThesisGeorgiaTech2012Impact,Rollett1991HighStrainRate}.

As shown in the present study, tensile bars with a characteristic width of a few millimetres can be cut out from a larger polycrystalline plate made of coarse-grained $\alpha$-uranium. The extruded area can be chosen such that only a small number of grains are present in the tensile bar aligned along the loading direction. High quality metallic surfaces, suitable for electron backscatter diffraction (EBSD) measurements, can be produced \citep{Sutcliffe2018Bristol} and the grain orientation can be found. In-situ digital image correlation (DIC) measurements are carried out to measure the axial strain in each grain and the lateral displacement of the tensile bar. The stress-strain curve of each grain is subsequently obtained and compared with the simulated curves obtained with the CPFE method. A nonlinear optimisation procedure, based on the Nelder-Mead algorithm \citep{NelderMead1965}, is used to find the CRSS and hardening rates of different slip systems. The optimisation procedure can validate the values of 4 parameters against data from a tensile bar containing 4 grains. Two tensile bars are used to validate the model by comparing the measured and simulated strain field. Different models for the hardening of the twin systems are compared \citep{RKALIDINDI2001837}.

The hardening of the most active slip system (``wall'' slip) in the coarse-grained material turns out to be lower than for fine-grained $\alpha$-uranium \citep{McCabeTome2010}. It will be shown that the hardening of the most active twin system $\{130\}\langle3\bar{1}0\rangle$ due to coplanar twins and ``floor'' slip is able to explain the experimental results, and are therefore important mechanisms. This is possible by choosing a grain in which one variant of this twin system has a Schmid factor greater than for the slip systems. The experimental technique and computational method developed in the present research can be applied to identify the CRSS and hardening parameters of other coarse grained materials.

In Section \ref{S:materialmodel} the crystal plasticity model for slip and twinning is described. Section \ref{S:twinsystemhardening2} contains the three different models {for hardening} of the twin systems which were tested. Experimental details are reported in section \ref{S:experimentaldetails}. Section \ref{S:simulationdetails} contains the simulation results and the optimisation procedure to find the CRSS and hardening rates. The discussion and conclusions then follow in sections \ref{S:discussion} and \ref{S:conclusions} respectively.


\section{Crystal plasticity finite element modelling}
\label{S:materialmodel}

The crystal plasticity framework is based on the decomposition of the deformation gradient $\bm{F}$ into an elastic part $\bm{F}_e$, describing the stretch of the crystal lattice, and a plastic part $\bm{F}_p$, describing slip and twinning \citep{ROTERS20123}:

\begin{equation}
\bm{F} = \bm{F}_e \bm{F}_p \ .
\label{eqn:elasticplasticdecomp}    
\end{equation}
The plastic deformation gradient $\bm{F}_p$ is an eigenstrain that does not contribute directly to the stress tensor. It has the following time evolution law \citep{KALIDINDI1998267}:

\begin{align}
\bm{L}_p = \dot{\bm{F}}_p \bm{F}_p^{-1} = & \left ( 1 - \sum_{\beta = 1}^{N_{\textrm{twin}}} f_\beta \right ) \sum_{\alpha=1}^{N_{\textrm{slip}}} \dot{\gamma}_{\alpha} \left ( \bm{\sigma} \right ) \bm{s}_\alpha \otimes \bm{n}_\alpha \nonumber\\
&+\sum_{\beta=1}^{N_{\textrm{twin}}} \dot{f}_\beta \left ( \bm{\sigma} \right ) \gamma^{\textrm{twin}}_{\beta} \bm{s}_\beta \otimes \bm{n}_\beta.
\label{eqn:plasticvelocitygrad}
\end{align}
The first term on the right hand side accounts for dislocation slip in the untwinned lattice, while the second term accounts for plastic deformation due to the creation of new twins. $\bm{L}_p$ is the plastic velocity gradient which relates $\bm{F}_p$ to its rate of change, $\dot{\bm{F}}_p$. $\dot{\gamma}_{\alpha} \left ( \bm{\sigma} \right )$ is the dislocation slip rate on slip system $\alpha$; defined by the slip direction, $\bm{s}_\alpha$, and slip plane normal $\bm{n}_\alpha$.  $f_\beta$ is the twin volume fraction of twin system $\beta$ (defined by the twin direction, $\bm{s}_{\beta}$, and twin plane normal $\bm{n}_{\beta}$) which increases at a rate of $\dot{f}_\beta \left ( \bm{\sigma} \right )$ whereas the shear produced by the twin system, $\gamma^{\textrm{twin}}_{\beta}$, is a constant. Note, that both $\dot{\gamma}_{\alpha} \left ( \bm{\sigma} \right )$ and $\dot{f}_\beta \left ( \bm{\sigma} \right )$ are stress dependent, as discussed in section \ref{S:twinsystemhardening}. The slip and twin systems used in the simulations are reported in table \ref{tab:sliptwinsystems}. 

The slip directions and slip plane normals are rotated from the lattice reference frame to the sample reference frame by a rotation matrix $\bm{R}$; which represents the grain orientation. $\bm{R}$ is updated at each time increment using the continuum elastic spin matrix, as explained in \citep{Clausmeyer2011}.

To model the nonlinear mechanical behaviour, at every time increment $\Delta t$ of the simulation, the Cauchy stress increment $\Delta \bm{\sigma} = \bm{\sigma} - \bm{\sigma}_0$ is calculated as a function of the deformation gradient, where $\bm{\sigma}_0$ and $\bm{\sigma}$ are the Cauchy stress at the beginning and end of the time increment respectively. This is given by the anisotropic Hooke's law:
\begin{equation}
\Delta \bm{\sigma} = \mathbb{C} \Delta \bm{\varepsilon}_\textrm{e} + \left ( \bm{W}_e \boldsymbol{\sigma}_0 - \boldsymbol{\sigma}_0 \bm{W}_e \right ) \Delta t \ ,
\label{eqn:cauchystressincr}
\end{equation}
where $\mathbb{C}$ is the fourth order elasticity tensor \citep{Fisher1958elastic} and $\bm{W}_e$ is the continuum elastic spin \citep{Belytschko2014}. The elastic constants for $\alpha$-uranium are reported in table \ref{tab:elasticconstants}. Details of the calculation of the elastic part of the small strain increment $\Delta \bm{\varepsilon}_\textrm{e}$ are reported in \citep{Grilli2019Uranium}. To find the stress increment that satisfies equations (\ref{eqn:elasticplasticdecomp}), (\ref{eqn:plasticvelocitygrad}) and (\ref{eqn:cauchystressincr}), an implicit equation for $\Delta \bm{\sigma}$ is solved using a Newton-Raphson algorithm, as detailed in \citep{DUNNE20071061,DunnePetrinicBook}. The crystal plasticity framework is implemented in a user material subroutine (UMAT) for the finite element software Abaqus, which solves the equilibrium equation \citep{TomiFionneAlan2013nonproportional}: $\nabla \cdot \bm{\sigma} = 0 $.
\begin{table*}[htb]
    \centering
    \begin{tabular}{|l|c|c|}
         \hline
         Slip system & $\bm{s}_{\alpha}^0$ & $\bm{n}_{\alpha}^0$ \\
         \hline
         $\alpha = 1$ (wall) & $\left [ 1 , 0 , 0 \right ]$ & $\left [ 0 , 1 , 0 \right ]$ \\
         $\alpha = 2$ (floor) & $\left [ 1 , 0 , 0 \right ]$ & $\left [ 0 , 0 , 1 \right ]$ \\
         $\alpha = 3$ (chimney) & $\left [ 0.437, -0.899 , 0 \right ]$ & $\left [ 0.899 , 0.437 , 0 \right ]$ \\
         $\alpha = 4$ (chimney) & $\left [ 0.437, 0.899 , 0 \right ]$ & $\left [ 0.899 , -0.437 , 0 \right ]$ \\
         $\alpha = 5$ (roof) & $\left [ 0.241, -0.495 , 0.835 \right ]$ & $\left [ 0.0 , 0.860 , 0.510 \right ]$ \\
         $\alpha = 6$ (roof) & $\left [ -0.241, -0.495 , 0.835 \right ]$ & $\left [ 0.0 , 0.860 , 0.510 \right ]$ \\
         $\alpha = 7$ (roof) & $\left [ 0.241, 0.495 , 0.835 \right ]$ & $\left [ 0.0 , 0.860 , -0.510 \right ]$ \\
         $\alpha = 8$ (roof) & $\left [ 0.241, -0.495 , -0.835 \right ]$ & $\left [ 0.0 , 0.860 , -0.510 \right ]$ \\
         \hline
         Twin system & $\bm{s}_{\beta}^0$ & $\bm{n}_{\beta}^0$ \\
         \hline
         $\beta = 1$ & $\left [ 0.825 , -0.565 , 0 \right ]$ & $\left [ 0.565 , 0.825 , 0 \right ]$ \\
         $\beta = 2$ & $\left [ -0.825 , -0.565 , 0 \right ]$ & $\left [ -0.565, 0.825 , 0 \right]$ \\
         \hline
    \end{tabular}
    \caption{Slip and twin systems used in the model \citep{McCabeTome2010}. The directions and plane normals are expressed in Cartesian coordinates and in the lattice coordinate system.}
    \label{tab:sliptwinsystems}
\end{table*}
\begin{table*}[!htb]
    \centering
    \begin{tabular}{|c|c|c|c|c|c|c|c|c|}
         \hline
         $\mathbb{C}_{11}$ & $\mathbb{C}_{12}$ & $\mathbb{C}_{13}$ & $\mathbb{C}_{22}$ & $\mathbb{C}_{23}$ & $\mathbb{C}_{33}$ & $\mathbb{C}_{44}$ & $\mathbb{C}_{55}$ & $\mathbb{C}_{66}$ \\
         \hline
         214.74 & 46.49 & 21.77 & 198.57 & 107.91 & 267.11 & 124.44 & 73.42 & 74.33 \\
         \hline
    \end{tabular}
    \caption{Elastic constants (GPa) at room temperature for the orthorhombic structure of $\alpha$-uranium \citep{Fisher1958elastic,BEELER2013143} in Voigt notation.}
    \label{tab:elasticconstants}
\end{table*}

\subsection{Constitutive model for slip and twinning}
\label{S:twinsystemhardening}

In this section the constitutive model to calculate the shear strain rate $\dot{\gamma}_{\alpha} \left ( \bm{\sigma} \right )$ for each slip system and the rate of increase of the twin volume fraction for each twin system is reported. A power law relationship is used for $\dot{\gamma}_{\alpha} \left ( \bm{\sigma} \right )$ \citep{ASARO1985923}:
\begin{equation}
\dot{\gamma}_{\alpha}(\bm{\sigma}) = \dot{\gamma}_0 \left | \frac{\tau_{\alpha}}{\tau_{\alpha}^{c}} \right |^n \textrm{sign} (\tau_{\alpha}) \ ,
\label{eqn:sliprulepowerlaw}
\end{equation}

where $\dot{\gamma}_0$ and $n$ are constants that determine the strain rate dependence and rate sensitivity of the slip activity. $\tau_{\alpha}$ is the resolved shear stress and $\tau_{\alpha}^{c}$ is the CRSS {of slip} system $\alpha$, which depends on the dislocation densities. Similarly, the rate of increase of the twin volume fraction depends on the resolved shear stress $\tau_{\beta}$ on the $\beta$ twin system and is also assumed to follow a power law \citep{AbdolvandDaymond2JMPS2013}: 
\begin{equation}
\gamma^{\textrm{twin}}_{\beta} \dot{f}_\beta \left ( \bm{\sigma} \right ) = 
\begin{cases}
\dot{\gamma}_0 \left | \frac{\tau_{\beta}}{\tau_{\beta}^{c}} \right |^n \ , \ \ \mbox{if} \ \tau_{\beta} > 0 \ , \\
0 \ , \ \ \ \ \ \ \ \ \ \ \ \mbox{if} \ \tau_{\beta} < 0 \ .
\end{cases}
\label{eqn:twinvolfracrate}
\end{equation}
The CRSS for twinning, $\tau_{\beta}^{c}$, is weakly temperature dependent \citep{CHRISTIAN19951} and it can depend on dislocation density \citep{AbdolvandDaymond2JMPS2013}. The present model is not able to resolve discrete twins, but only the average volume fraction can be studied. $\tau_{\beta}^{c}$ has to be interpreted as the stress necessary to nucleate individual twins \citep{REMY1978443}.
The stress necessary for twin propagation \citep{QIAO201670} and migration \citep{Ojha2014} is typically lower than this value. These softening mechanisms, leading to stress drops \citep{PATRIARCA2013165}, are not included in the present model. 

A twin system can be activated only by a positive resolved shear stress \citep{KALIDINDI1998267}. The CRSS for slip is linked to the evolution of dislocation densities \citep{BEYERLEIN2008867,Mecking19811865,Madec2002}:

\begin{equation}
\tau_{\alpha}^{c} = \tau_{\alpha}^{0} + 0.9 \ b_\alpha \mu_\alpha \sqrt{\rho_{\alpha}^{\textrm{for}}} - 0.086 \ b_\alpha \mu_\alpha \sqrt{\rho^\textrm{sub}} \log \left ({b_\alpha \sqrt{\rho^\textrm{sub}}} \right ) \ ,
\label{eqn:crsstaylorlaw}
\end{equation}
where $\tau_{\alpha}^{0}$ is the initial slip resistance, $b_\alpha$ is the Burgers vector and $\mu_\alpha$ is the projected shear modulus \citep{Grilli2019Uranium} of the $\alpha$ slip system. $\rho_{\alpha}^{\textrm{for}}$ is the forest dislocation density, while $\rho^\textrm{sub}$ represents dislocation debris that forms stable dislocation substructures at the onset of stage IV hardening \citep{BEYERLEIN2008867,Madec2002}. The logarithmic term in (\ref{eqn:crsstaylorlaw}) represents the stress for gliding dislocations to bow out through arrays of locked dislocations {with spacing} $1/\sqrt{\rho^\textrm{sub}}$. This model was originally developed by \cite{BEYERLEIN2008867} for pure Zr, who provide further details. 

The hardening rate is determined by the time evolution of the dislocation densities \citep{McCabeTome2010}:
\begin{align}
\label{eqn:rhofor}
\dot{\rho}_{\alpha}^\textrm{for} &= k_{\alpha}^1 \frac{\left| \dot{\gamma}_{\alpha} (\boldsymbol{\sigma}) \right|}{b_\alpha} \Bigg[ \sqrt{\rho_{\alpha}^{\textrm{for}}} - \hat{d}_{\alpha}  \rho_{\alpha}^{\textrm{for}} \Bigg] \ , \\
\label{eqn:rhosub}
\dot{\rho}^\textrm{sub} &= 1800 k_1^1 \hat{d}_{1}
\rho_{1}^\textrm{for} \sqrt{\rho^\textrm{sub}} \left | \dot{\gamma}_{1} (\boldsymbol{\sigma}) \right | \ ,
\end{align}
where $k_{\alpha}^1$ are dimensionless parameters and $\hat{d}_{\alpha}$ is the annihilation distance for slip system $\alpha$. 

{The initial slip resistances $\tau_{\alpha}^{0}$ for $\alpha=1,2,3$ (``wall'', ``floor'', ``chimney'' slip systems) are three of the four parameters used in the optimisation procedure for the CRSS, described in section} \ref{S:optimization}, {and the dimensionless parameters $k_{\alpha}^1$ are three of the four parameters used in the optimisation procedure for the hardening.}

\subsection{Constitutive model for twin systems hardening}
\label{S:twinsystemhardening2}

In the present study three hardening models for the twin systems are compared. The CRSS of the twin systems $(130)[3\bar{1}0]$ and $(\bar{1}30)[\bar{3}\bar{1}0]$ will be indicated by $\tau_{(130)}^{c}$ and $\tau_{(\bar{1}30)}^{c}$ in the following. The corresponding twin volume fractions will be indicated by $f_{(130)}$ and $f_{(\bar{1}30)}$. The first hardening model takes into account the interaction between twin and slip systems \citep{AbdolvandDaymond2JMPS2013}. The CRSS for twinning is approximated with a linear function of the forest dislocation densities: %
\begin{flalign}
\text{Twin-slip:} && \tau_{\beta}^{c} &= \tau_{\beta}^{0} + k_\beta^1 \mu_\beta b_\beta \sum_{\alpha=1}^{N_\textrm{slip}} b_\alpha \rho_\alpha^{\textrm{for}} \ . &
\label{eqn:twinsliphardening}
\end{flalign}

The second hardening model \citep{ROTERS20101152} takes into account the interaction between the noncoplanar twin systems $(130)[3\bar{1}0]$ and $(\bar{1}30)[\bar{3}\bar{1}0]$:
\begin{flalign}
\label{eqn:twintwinhardening}
\text{Twin noncoplanar:} && \tau_{(130)}^{c} &= \tau_{\beta}^{0} + k_\beta^1 \mu_\beta f_{(\bar{1}30)} \ , & \\
\label{eqn:twintwinhardening2}
&& \tau_{(\bar{1}30)}^{c} &= \tau_{\beta}^0 + k_\beta^1 \mu_\beta f_{(130)} \ . &
\end{flalign}

The third hardening model \citep{ROTERS20101152} also includes the interaction between coplanar twin systems:
\begin{flalign}
\text{Twin coplanar:} && \tau_\beta^c &= \tau_{\beta}^{0} + k_\beta^1 \mu_\beta \left ( f_{(130)} + f_{(\bar{1}30)} \right ) \ . &
\label{eqn:twinselfhardening}
\end{flalign}
$\tau_{\beta}^{0}$ is the initial twin resistance, $\mu_\beta$ is the projected shear modulus \citep{Grilli2019Uranium} on the twin system and $k_\beta^1$ is a dimensionless interaction parameter. $\tau_{\beta}^{0}$ is the fourth parameter used in the optimisation procedure for the CRSS, described in section \ref{S:optimization}. $k_\beta^1$ is the fourth parameter used in the optimisation procedure for the hardening. The three hardening models are referred to as the twin-slip, twin-noncoplanar and twin-coplanar model from here on. The complete set of model parameters are reported in Table \ref{tab:modelparameters}.
\begin{table*}[!htb]
    \centering
    \begin{tabular}{|l|c|}
         \hline
         Slip law parameters  & Eqn. \eqref{eqn:sliprulepowerlaw} \\
         \hline
         Plastic strain rate coefficient ($\dot{\gamma}_0$) & 0.001 s$^{-1}$ \\
         \hline
         Plastic strain rate exponent ($n$) \citep{IrastorzaLanda2017} & 20 \\
         \hline\hline
         Twin law parameters  & Eqn. \eqref{eqn:twinvolfracrate} \\
         \hline
         Magnitude of shear due to twinning ($\gamma^{\textrm{twin}}_{\beta}$) \citep{CAHN195349} & 0.299 \\
         \hline
         Initial twin volume fraction $f_\beta \left ( t=0 \right )$ & 0 \\
         \hline\hline
         Slip hardening law parameters  & Eqn. \eqref{eqn:crsstaylorlaw} \\
         \hline
         Constant friction stress (roof slip) ($\tau_{5}^0$;$\tau_{6}^0$;$\tau_{7}^0$;$\tau_{8}^0$) \citep{CalhounAgnew2013ThermalStrains} & 235 MPa \\
         \hline
         Burgers vector (wall slip) ($b_{1}$) \citep{Lukesh1949} & 0.285 nm \\
         \hline
         Burgers vector (floor slip) ($b_{2}$) \citep{Lukesh1949} & 0.285 nm \\
         \hline
         Burgers vector (chimney slip) ($b_{3}$;$b_{4}$) \citep{Lukesh1949} & 0.651 nm \\
         \hline
         Burgers vector (roof slip) ($b_{5}$;$b_{6}$;$b_{7}$;$b_{8}$) \citep{Lukesh1949} & 1.185 nm \\
         \hline
         Projected shear modulus (wall slip) ($\mu_1$) \citep{Fisher1958elastic} & 74.330 GPa \\
         \hline
         Projected shear modulus (floor slip) ($\mu_2$) \citep{Fisher1958elastic} & 73.420 GPa \\
         \hline
         Projected shear modulus (chimney slip) ($\mu_3,\mu_4$) \citep{Fisher1958elastic} & 92.255 GPa \\
         \hline
         Projected shear modulus (roof slip) ($\mu_5,\mu_6$,$\mu_7,\mu_8$) \citep{Fisher1958elastic} & 115.67 GPa \\
         \hline\hline
         Dislocation density evolution law parameters  & Eqns. \eqref{eqn:rhofor},\eqref{eqn:rhosub} \\
         \hline
         Dislocation multiplication prefactor (roof slip) ($k_5^1$;$k_6^1$;$k_7^1$;$k_8^1$) \citep{McCabeTome2010} & 0.948 \\
         \hline
         Dislocation annihilation length (wall slip) ($\hat{d}_{1}$) \citep{McCabeTome2010} & 0.936 $\mu$m \\
         \hline
         Dislocation annihilation length (floor slip) ($\hat{d}_{2}$) \citep{McCabeTome2010} & 0.429 $\mu$m \\
         \hline
         Dislocation annihilation length (chimney slip) ($\hat{d}_{3}$,$\hat{d}_{4}$) \citep{McCabeTome2010} & 0.174 $\mu$m \\
         \hline
         Dislocation annihilation length (roof slip) ($\hat{d}_{5}$,$\hat{d}_{6}$,$\hat{d}_{7}$,$\hat{d}_{8}$) \citep{McCabeTome2010} & 0.124 $\mu$m \\
         \hline
         Initial dislocation density $\rho_{\alpha}^\textrm{for} \left ( t=0 \right )=\rho^\textrm{sub} \left ( t=0 \right )$ & $10^{10}$ m$^{-2}$ \\
         \hline\hline
         Twin hardening law parameters  & Eqns. \eqref{eqn:twinsliphardening}-\eqref{eqn:twinselfhardening} \\
         \hline
         Twin Burgers vector ($b_{\beta}$) & 0.1036 nm \\
         \hline
         Twin system projected shear modulus ($\mu_\beta$) & 99.537 GPa \\
         \hline
         Initial twin resistance (fitting parameter)
         $\tau_{\beta}^{0}$ & Table \ref{tab:optimizedparameters} \\
         \hline
         Hardening coefficients (fitting parameters) $k_\beta^1$ & Table \ref{tab:optimizedparameters} \\
         \hline
    \end{tabular}
    \caption{Model parameters.}
    \label{tab:modelparameters}
\end{table*}

\section{Experimental details}
\label{S:experimentaldetails}

Experiments were performed to measure the full-field surface strains during tensile deformation of cast $\alpha$-uranium specimens with a known microstructure, using digital image correlation (DIC). Two dog-bone specimens were manufactured by AWE plc via wire electro-discharge machining (EDM) from a cast plate. The gauge dimensions were 32.00\,mm $\times$ 6.00\,mm, with a specimen thickness of 1.50\,mm.

\subsection{Sample preparation and EBSD}
\label{S:sampleprep}

Sample preparation and EBSD microstructural characterisation were performed at the Interface Analysis Centre (IAC), School of Physics, University of Bristol, UK. Each specimen was mounted in Struers ClaroCit resin, and mechanically ground with 120 grit SiC paper to remove the oxide layer. Grinding with progressively finer papers, from 320 to 2500 grit, was followed by polishing with 6\,$\mu$m and 3\,$\mu$m diamond suspension. The specimens were removed from the resin with acetone, and electropolished at a voltage of 12\,V in an electrolyte comprising a 10:6:6 volumetric ratio of ethanol, ethylene glycol, and phosphoric acid. The electrolyte was continuously stirred and polishing proceeded for between 10 and 20 minutes, until a smooth, reflective surface was observed.

EBSD analysis was performed on a Zeiss Sigma HD VP Field Emission SEM, with EDAX EBSD camera and associated OIM software. A series of large-area maps covering the whole gauge area were recorded with a slight overlap between each scan region. The EBSD datasets were post-processed using the MTEX toolbox \citep{Bachmann2011} in MATLAB. The individual orientation maps were stitched together to produce a single map for the gauge region of each specimen.

\begin{figure}[!htb]
\includegraphics[width=\textwidth]{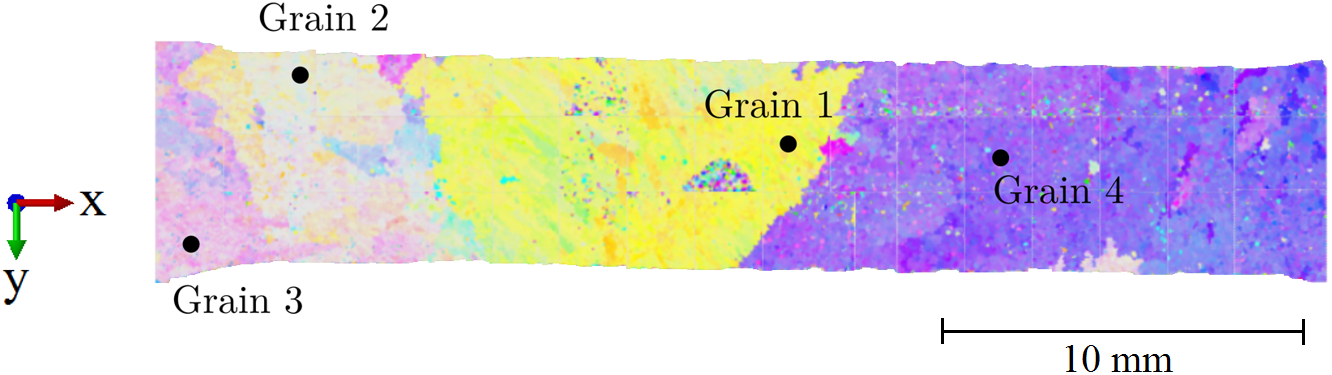}
\caption{\label{fig:TB3EBSD} EBSD orientation map of the first tensile bar. The black dots indicate the crystal orientations chosen for the four grains.}
\end{figure}

\begin{figure}[!htb]
\includegraphics[width=\textwidth]{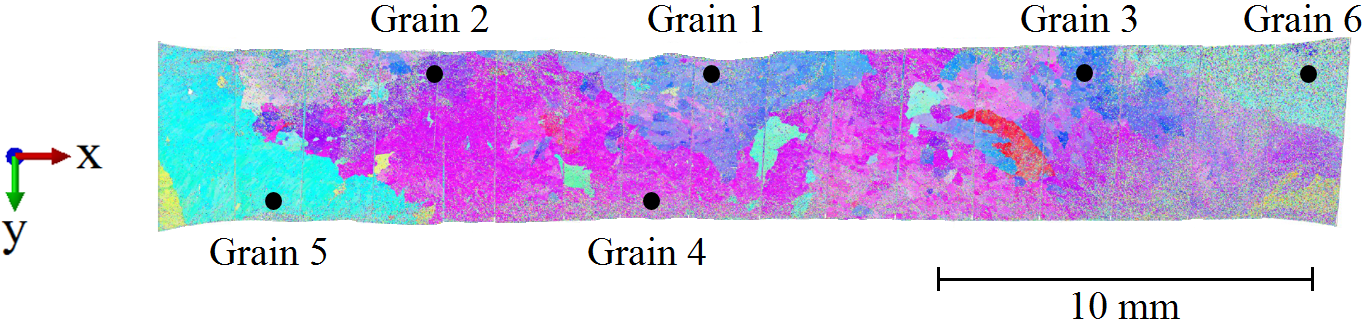}
\caption{\label{fig:TB1EBSD} EBSD orientation map of the second tensile bar. The black dots indicate the crystal orientations chosen for the six grains.}
\end{figure}

Figures \ref{fig:TB3EBSD} and \ref{fig:TB1EBSD} show orientation maps from the two tensile bars, with colours representing the crystal direction parallel to the z direction (into-page). The microstructure is coarse, with sub-grains in excess of 200\,$\mu$m in diameter clustered together into regions of similar orientations. Grain clusters are labelled according to their orientation; with 4 grains in the first tensile bar and 6 in the second.

\subsection{In-situ mechanical testing}
\label{S:insitutest}

A speckle pattern of black and white acrylic paint was applied to the surface of each specimen with an airbrush. Tensile testing was performed on a Shimadzu Autograph AGS-X universal test machine with a 5\,kN load cell, connected to a data acquisition PC running Shimadzu Trapezium-X software. 

The specimen was enclosed in a bespoke sample holder to prevent contamination of the tensile grips. A window in the front face of the holder allows viewing of the specimen surface with the DIC cameras. The specimen was loaded in parallel with a thin ductile copper bar which would absorb energy in the event of specimen failure to prevent loss of radioactive material. The mechanical properties of the copper {were} characterised before the test; its flow curve was used to calculate the stress in the $\alpha$-uranium specimen from the total load measured by the load cell. Each specimen was loaded in displacement control at a rate of either 0.05\,mm/min or 0.025\,mm/min.

DIC was performed with a GOM ARAMIS 5M stereo DIC system. The camera mounting frame was rotated by 90$^{\circ}$ from its usual horizontal orientation such that the affixed cameras were positioned above one another and angled inwards towards the specimen {at} an angle of 12.5$^{\circ}$. This allowed the gauge section of the specimen to be viewed in a portrait orientation. A calibration process allowed the co-ordinates in each image to be related to 3D co-ordinates in a small calibrated volume centred on the specimen surface. Two camera DIC systems as used here allow the separation of in-plane strains from out-of-plane displacements, which cause strain artifacts in single-camera DIC. Illumination was provided by polarised light from high-intensity LEDs. A pair of DIC images was automatically acquired in the ARAMIS software every 2.5 seconds during the test.

The DIC analysis was performed with a facet size of 20$\times$20 pixels and a 50\% overlap between each facet. A rigid-body movement of the specimen before testing was used to calculate the in-plane strain error as $\pm$0.1\%, and the error in out-of-plane displacement as $\pm$30\,$\mu$m. The boundaries between the grain clusters in figure \ref{fig:TB3EBSD} were used to segment the DIC data such that the mean strain within each grain cluster can be calculated. This enables the plotting of stress-strain curves for the entire gauge volume (figure \ref{fig:gaugestressstrain}) and for the individual grain clusters (figure \ref{fig:grain1stressstrain}), for comparison with the simulation results.

\section{Simulation details and results}
\label{S:simulationdetails}

\begin{figure}[!htb]
\centering
\subfloat[(a)]{\includegraphics[width=0.39\textwidth]{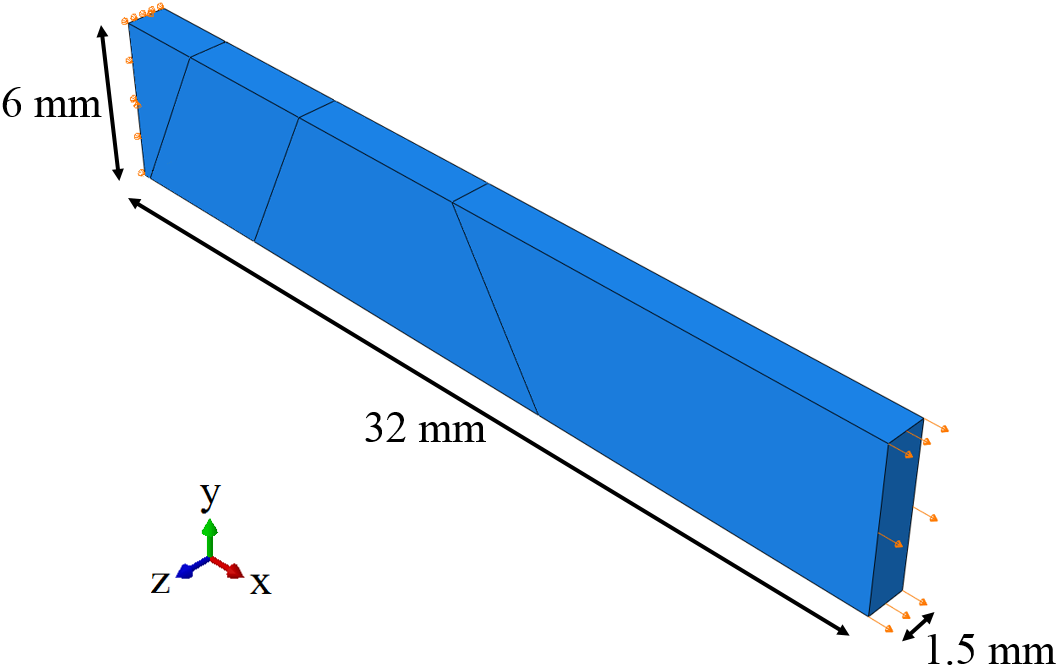}}
\subfloat[(b)]{\includegraphics[width=0.32\textwidth]{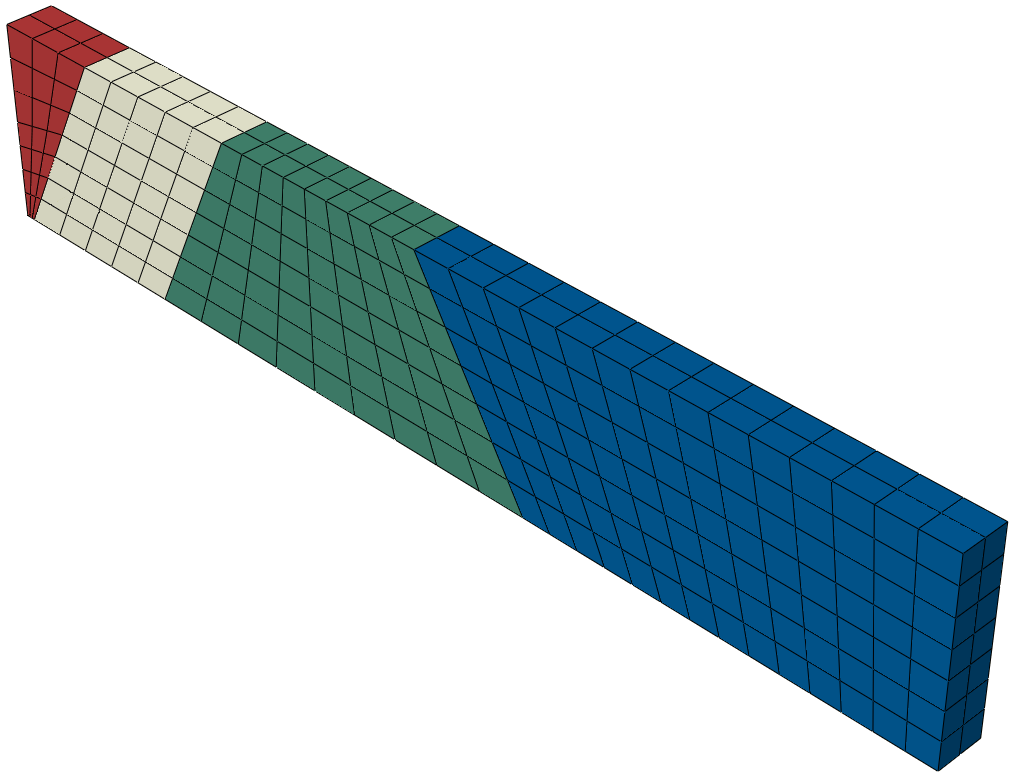}}
\subfloat[(c)]{\includegraphics[width=0.32\textwidth]{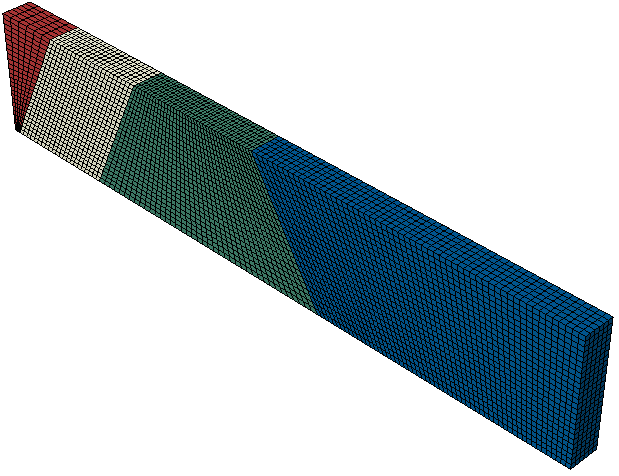}}
\caption{\label{fig:meshbcsmall}(a) Representative volume of the first tensile bar. (b) Coarse mesh used during the optimisation procedure; different colours correspond to the Four different grains. (c) Fine mesh used for the validation procedure.}
\end{figure}

The representative volume used for the first tensile bar is shown in figure \ref{fig:meshbcsmall} (a). The bar was 32 mm long and 6 mm wide. Displacement along the x axis was applied on the surface x = 32 mm. Zero displacement along the x direction was imposed on the surface x = 0. The point (x,y,z) = (0, 3, 0.75) mm was fixed to prevent translation of the surface x = 0. The points (x,y,z) = (0, 3, 1.5) mm and (x,y,z) = (0, 6, 0.75) mm can translate only along the z and y directions. This prevents rotation around the x axis of the surface x = 0. The texture of the representative volume consisted of four grains with different orientations (figure \ref{fig:meshbcsmall}). This is a simplification compared to the experimental EBSD map (figure \ref{fig:TB3EBSD}) but is useful to maintain constant Schmid factors in each grain and, therefore, to better understand the slip-twin activity in each grain. This approximation also avoids the introduction of spurious data at points where the experimental EBSD has larger uncertainty. An uncertainty quantification on the grain orientations has been carried out and its effect on the computational results is discussed in section \ref{S:discussion}.

The coarser mesh used for the optimisation procedure is shown in figure \ref{fig:meshbcsmall} (b). It consisted of 434 elements with an average size of around 1 mm. The coarser mesh is necessary to reduce the computational time during the optimisation procedure, which requires hundreds of simulations. A finer mesh, of 19500 elements, shown in figure \ref{fig:meshbcsmall} (c), was used to repeat the simulation with the optimised parameters and compare the simulated and experimental strain fields.  

The maximum time increment used in the simulations was 0.04 s and the total time was 20 s, after which a total strain of 2\% was reached. Therefore, the simulated strain rate was $10^{-3}$ s$^{-1}$: this value corresponds to the constant $\dot{\gamma}_0$ in equation \ref{eqn:sliprulepowerlaw} and was chosen to simulate the regime in which the model is strain rate independent \citep{CAPOLUNGO200942}. In this regime, the resolved shear stress on a slip or twin system remains close to the CRSS $\tau_{\alpha}^{c}$ or $\tau_{\beta}^{c}$ respectively. At the beginning of the simulations, the twin volume fractions were set to zero and the dislocation densities were set to $10^{10}$ m$^{-2}$.

\subsection{Optimisation procedure}
\label{S:optimization}

The first optimisation procedure was carried out to find the initial CRSS parameters $\tau_\alpha^0$ and $\tau_\beta^0$ for ``wall'' slip ($\alpha = 1$), ``floor'' slip ($\alpha = 2$), ``chimney'' slip ($\alpha = 3$) and {the} $\{130\}\langle3\bar{1}0\rangle$ twin ($\beta = 1$). A nonlinear optimisation procedure, based on the Nelder-Mead algorithm \citep{NelderMead1965}, was used, which is implemented in the {\texttt{scipy.optimize} package} \citep{SciPy}. This algorithm was used because it does not require an analytical expression for the Jacobian. At every iteration of the algorithm, the simulation with the coarse mesh in figure \ref{fig:meshbcsmall} (b) was run with a set of parameters $\tau_\alpha^0$, $\tau_\beta^0$. For each grain in figure \ref{fig:TB3EBSD} the axial strain was averaged; denoted $\bar{\varepsilon}^\textrm{G1}$, $\bar{\varepsilon}^\textrm{G2\&3}$ and $\bar{\varepsilon}^\textrm{G4}$, for grains 1, 2 and 3, 4 respectively. The strain was averaged together in grains 2 and 3 because their mechanical behaviour was similar. The total strain in the tensile bar is referred to as $\bar{\varepsilon}^\textrm{gauge}$ and the stress was averaged over the load surface in figure \ref{fig:meshbcsmall} (a). This post-processing allows stress-strain curves to be extracted from the tensile bar and from the individual grains. The yield stresses of the tensile bar, $\sigma_y^\textrm{gauge}$, and of the individual grains, $\sigma_y^\textrm{G1}$, $\sigma_y^\textrm{G2\&3}$, $\sigma_y^\textrm{G4}$, were defined using the 0.1\% offset criterion \citep{PHAM2013143} for both the experiment and the simulation. The residual minimised by the Nelder-Mead algorithm was:
\begin{equation}
R_\textrm{CRSS} = \left ( \sigma_{y,\textrm{exp}}^\textrm{gauge} - \sigma_{y,\textrm{sim}}^\textrm{gauge} \right )^2 + \left ( \sigma_{y,\textrm{exp}}^\textrm{G1} - \sigma_{y,\textrm{sim}}^\textrm{G1} \right )^2 + \left ( \sigma_{y,\textrm{exp}}^\textrm{G2\&3} - \sigma_{y,\textrm{sim}}^\textrm{G2\&3} \right )^2 + \left ( \sigma_{y,\textrm{exp}}^\textrm{G4} - \sigma_{y,\textrm{sim}}^\textrm{G4} \right )^2 \ ,
\label{eqn:residualcrss}
\end{equation}
where the subscripts exp and sim indicate experimental and simulated quantities. After every iteration of the Nelder-Mead algorithm, the parameters $\tau_\alpha^0$ and $\tau_\beta^0$ are updated. This first optimisation procedure is carried out without considering the hardening of the twin systems because this does not significantly affect the yield stress. The CRSS parameters $\tau_\alpha^0$ and $\tau_\beta^0$ that minimise the residual $R_\textrm{CRSS}$ in equation (\ref{eqn:residualcrss}) are given in table \ref{tab:optimizedparameters}.
\begin{table*}[!htb]
    \begin{tabular}{|l|c|}
         \hline
         CRSS parameters optimisation & Eqn. \eqref{eqn:crsstaylorlaw} and \eqref{eqn:twinsliphardening}-\eqref{eqn:twinselfhardening} \\
         \hline
         Constant friction stress (wall slip) ($\tau_1^0$) & 24.5 MPa \\
         \hline
         Constant friction stress (floor slip) ($\tau_2^0$) & 85.5 MPa \\
         \hline
         Constant friction stress (chimney slip) ($\tau_3^0$;$\tau_4^0$) & 166.5 MPa \\
         \hline
         Twin resistance (\{130\} twin system) ($\tau_\beta^0$) & 55.5 MPa \\
         \hline
    \end{tabular}
    \begin{tabular}{|l|c|c|c|}
         \hline
         Hardening parameters optimisation & Eqn. \eqref{eqn:twinsliphardening} & \eqref{eqn:twintwinhardening}-\eqref{eqn:twintwinhardening2} & \eqref{eqn:twinselfhardening} \\
         \hline
         Dislocation multiplication prefactor (wall slip) ($k_{1}^1$) & 0.0121 & 0.0088 & 0.0123 \\
         \hline
         Dislocation multiplication prefactor (floor slip) ($k_{2}^1$) & 0.36 & 1.69 & 1.80 \\
         \hline
         Dislocation multiplication prefactor (chimney slip) ($k_{3}^1$;$k_{4}^1$) & 0.136 & 0.176 & 0.0 \\
         \hline
         Twin system hardening parameter ($k_\beta^1$) & 4706 & 0.0570 & 0.00097 \\
         \hline
    \end{tabular}
    \caption{Parameters obtained from the optimisation procedures for the CRSS and for the hardening rate. The three values of the parameters $k_{1}^1$, $k_{2}^1$, $k_{3}^1$, $k_{4}^1$ and $k_\beta^1$ are obtained using the twin-slip, twin-noncoplanar and twin-coplanar model respectively.}
    \label{tab:optimizedparameters}
\end{table*}

A second optimisation procedure was then carried out to find the hardening parameters $k_\alpha^1$ and $k_\beta^1$ for ``wall'' slip ($\alpha = 1$), ``floor'' slip ($\alpha = 2$), ``chimney'' slip ($\alpha = 3$) and {the} $\{130\}\langle3\bar{1}0\rangle$ twin ($\beta = 1$) using the CRSS parameters $\tau_\alpha^0$ and $\tau_\beta^0$ obtained from the first optimisation procedure. The hardening parameters $k_\alpha^1$ and $k_\beta^1$ were updated after every iteration of the Nelder-Mead optimisation algorithm, in which a simulation with the coarse mesh in figure \ref{fig:meshbcsmall} (b) up to 2\% total strain was run. The maximum stress of the tensile bar reached during the deformation is denoted $\sigma^\textrm{gauge}$. The maximum total axial strain in the tensile bar is denoted $\bar{\varepsilon}^\textrm{gauge}$, and the maximum average axial strains reached during the deformation in the individual grains, $\bar{\varepsilon}^\textrm{G1}$, $\bar{\varepsilon}^\textrm{G2\&3}$, $\bar{\varepsilon}^\textrm{G4}$, were calculated after each simulation. The residual minimised during the second optimisation {procedure:} 
\begin{equation}
R_\textrm{hard} =  R_\textrm{hard}^\sigma + \left ( 10^4 \ \textrm{MPa} \right )^2  R_\textrm{hard}^{\bar{\varepsilon}} \ ,
\label{eqn:residualhardening}
\end{equation}
where
\begin{equation}
R_\textrm{hard}^\sigma = 4 \left ( \sigma_{\textrm{exp}}^\textrm{gauge} - \sigma_{\textrm{sim}}^\textrm{gauge} \right )^2 \ ,
\label{eqn:residualhardeningstress}
\end{equation}
and
\begin{equation}
R_\textrm{hard}^{\bar{\varepsilon}} = \left ( \bar{\varepsilon}_{\textrm{exp}}^\textrm{gauge} - \bar{\varepsilon}_{\textrm{sim}}^\textrm{gauge} \right )^2 + \left ( \bar{\varepsilon}_{\textrm{exp}}^\textrm{G1} - \bar{\varepsilon}_{\textrm{sim}}^\textrm{G1} \right )^2 + \left ( \bar{\varepsilon}_{\textrm{exp}}^\textrm{G2\&3} - \bar{\varepsilon}_{\textrm{sim}}^\textrm{G2\&3} \right )^2 + \left ( \bar{\varepsilon}_{\textrm{exp}}^\textrm{G4} - \bar{\varepsilon}_{\textrm{sim}}^\textrm{G4} \right )^2 \ ,
\label{eqn:residualhardeningstrain}
\end{equation}
was {determined} for the three different hardening models for the twin systems reported in section \ref{S:twinsystemhardening2}. The hardening parameters $k_\alpha^1$ and $k_\beta^1$ that minimise the residual $R_\textrm{hard}$ in (\ref{eqn:residualhardening}) are reported in table \ref{tab:optimizedparameters}.

\begin{figure}[!htb]
\centering
\subfloat[(a)]{\includegraphics[height=0.37\textwidth]{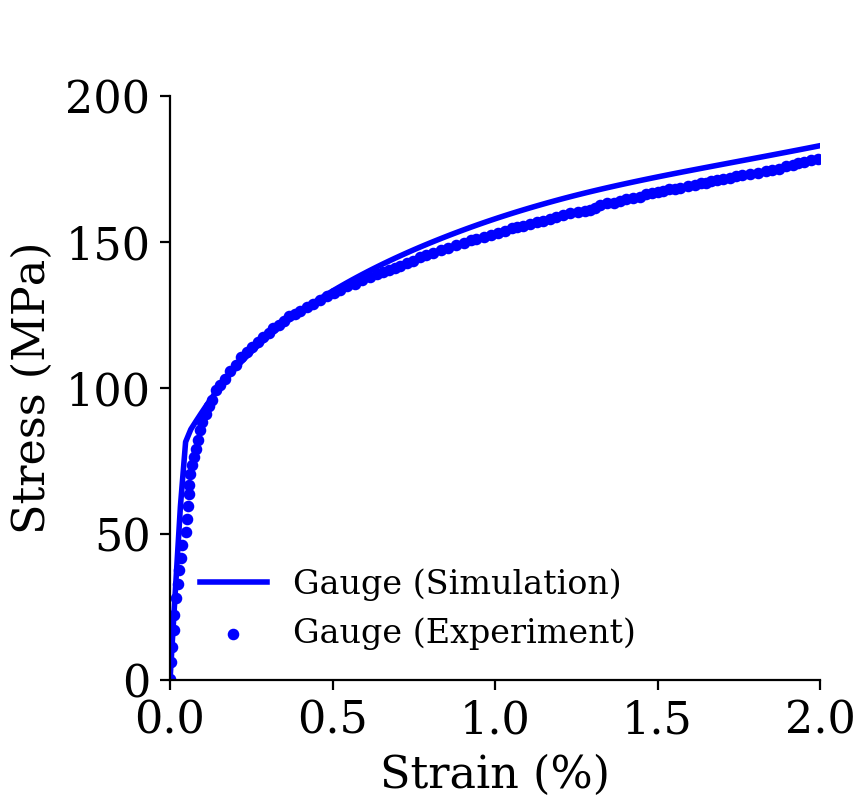}}
\subfloat[(b)]{\includegraphics[height=0.37\textwidth]{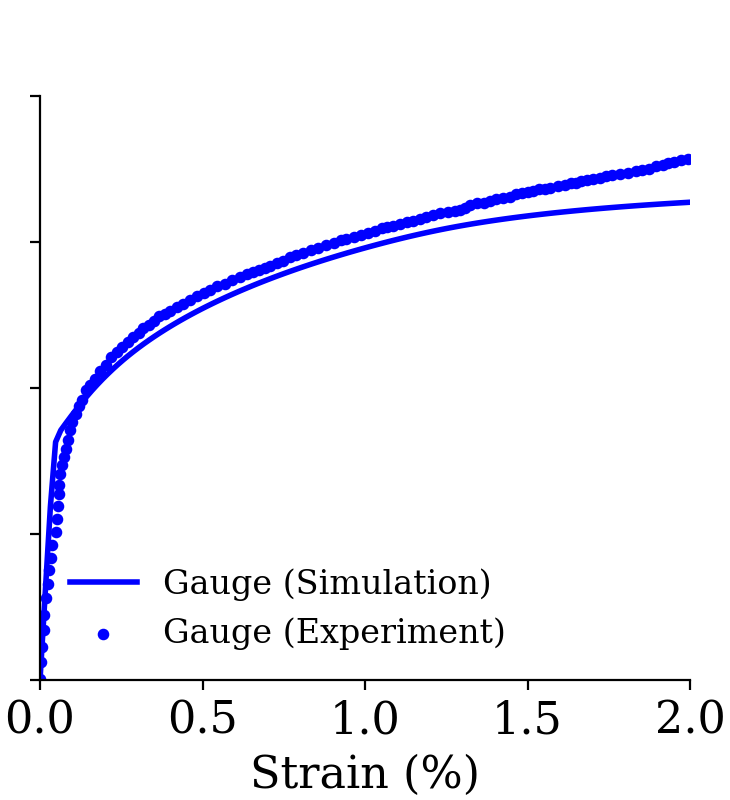}}
\subfloat[(c)]{\includegraphics[height=0.37\textwidth]{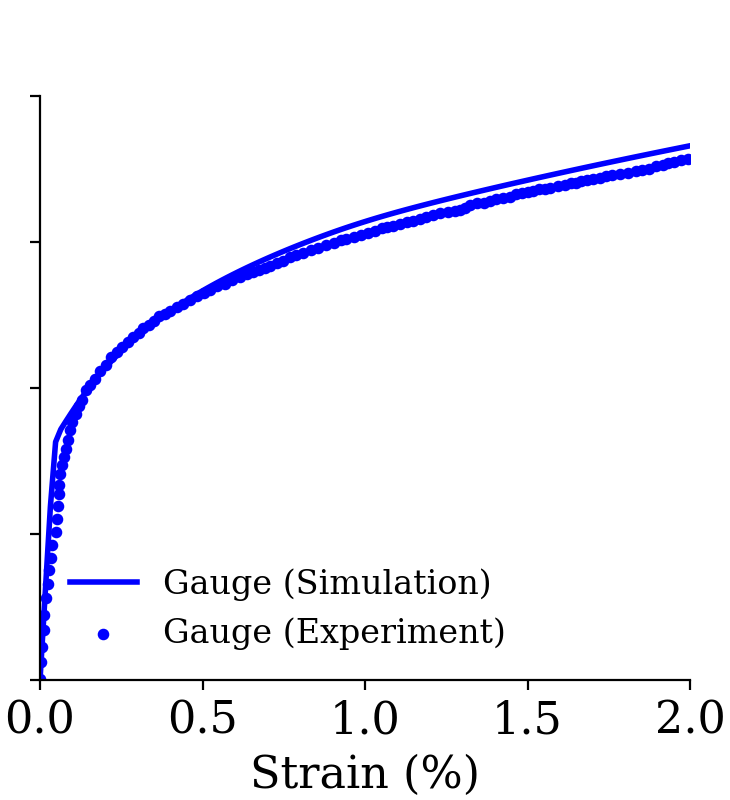}}
\caption{\label{fig:gaugestressstrain} Simulated and measured gauge stress-strain curves after the optimisation procedure using (a) the twin-slip model; equation  (\ref{eqn:twinsliphardening}), (b) the twin-noncoplanar model; equations (\ref{eqn:twintwinhardening})-(\ref{eqn:twintwinhardening2}) and (c) twin-coplanar model; equation (\ref{eqn:twinselfhardening}).}
\end{figure}

\begin{figure}[!htb]
\centering
\subfloat[(a)]{\includegraphics[height=0.33\textwidth]{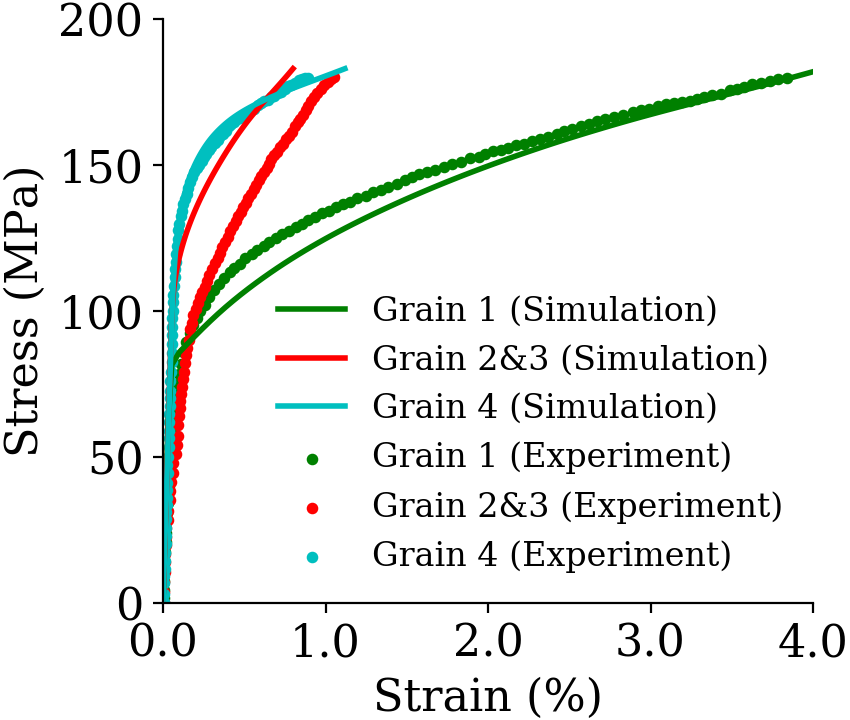}}
\subfloat[(b)]{\includegraphics[height=0.33\textwidth]{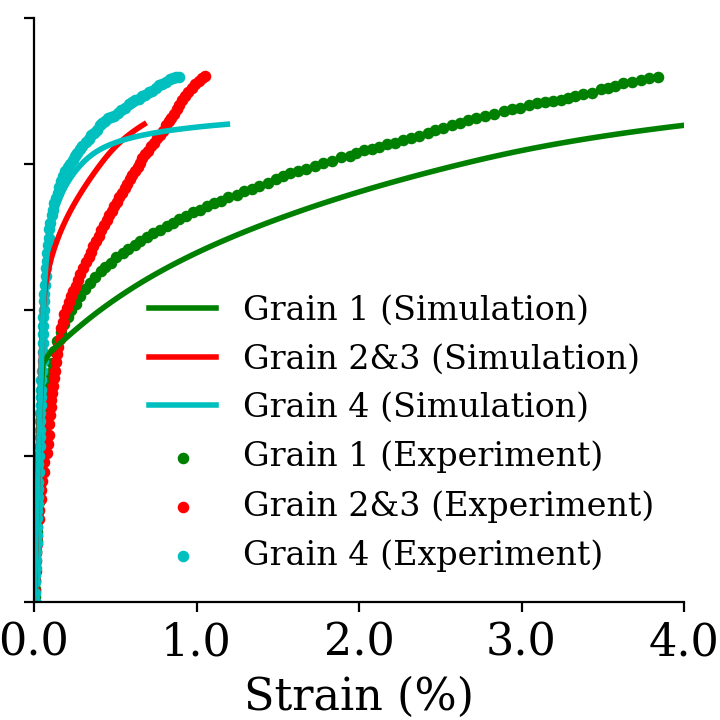}}
\subfloat[(c)]{\includegraphics[height=0.33\textwidth]{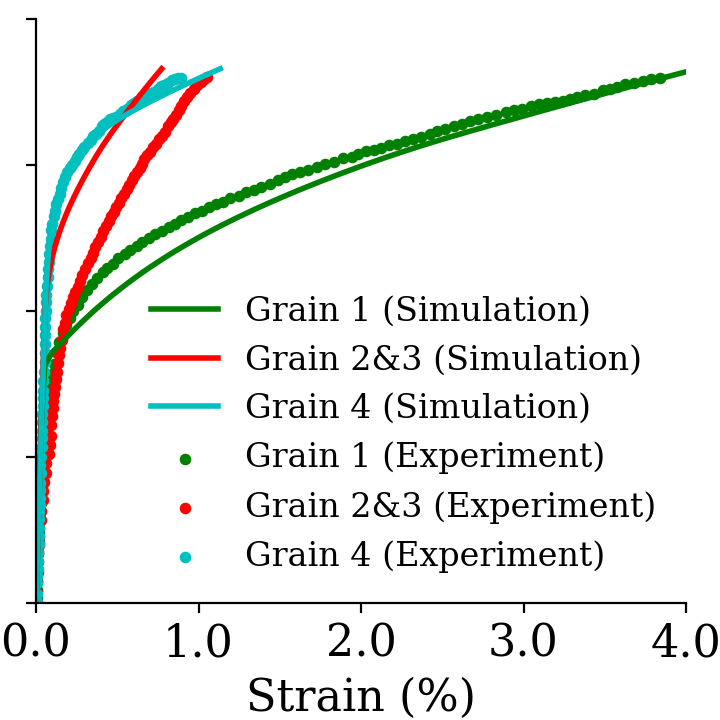}}
\caption{\label{fig:grain1stressstrain} Simulated and measured stress-strain curves in grain 1, grain 2\&3 and grain 4 after the optimisation procedure using (a) the twin-slip model; equation  (\ref{eqn:twinsliphardening}), (b) the twin-noncoplanar model; equations (\ref{eqn:twintwinhardening})-(\ref{eqn:twintwinhardening2}) and (c) twin-coplanar model; equation (\ref{eqn:twinselfhardening}).}
\end{figure}

The simulated stress-strain curves obtained using the optimised parameters and the three different hardening models for the twin systems are shown in figure \ref{fig:gaugestressstrain} alongside the experimental data. 
Figure \ref{fig:grain1stressstrain} shows stress as a function of the average axial strains $\bar{\varepsilon}^\textrm{G1}$, $\bar{\varepsilon}^\textrm{G2\&3}$, $\bar{\varepsilon}^\textrm{G4}$ in the individual grains for the three different hardening models for the twin systems. The stress-strain curves agree with the experiment for the twin-slip (figure \ref{fig:grain1stressstrain} (a)) and for the twin-coplanar (figure \ref{fig:grain1stressstrain} (c)) interaction models. In grains 2 and 3, the maximum stress agrees with the experiment, as shown in figures \ref{fig:grain1stressstrain} (a) and (c), while the maximum strain is underestimated by the models.


The twin-noncoplanar interaction model is not able to match the stress value reached in the experiment, as shown in figures \ref{fig:gaugestressstrain} (b) and \ref{fig:grain1stressstrain} (b). It overestimates the strain accommodated in grain 4, as shown in figure \ref{fig:grain1stressstrain} (b). This will be interpreted in the following section according to the activity of the slip and twin systems.



\subsection{Fine mesh simulation results {and model validation}}
\label{S:finemeshsimulations}

\begin{figure}[!htb]
\centering
\includegraphics[width=1.0\textwidth]{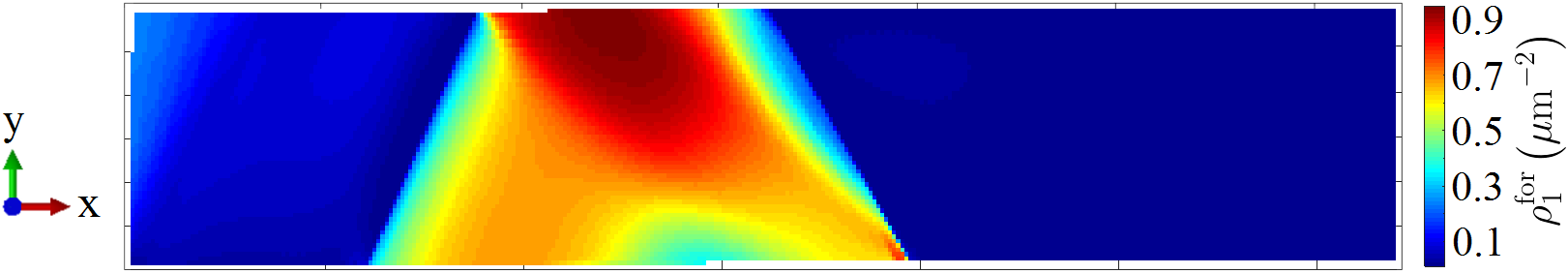}
\caption{\label{fig:rhowallbar3}Dislocation density $\rho_{1}^\textrm{for}$ in the wall slip system in the first tensile bar at 1\% total strain {using the twin-coplanar model of equation} (\ref{eqn:twinselfhardening}).}
\end{figure}

Once the optimal parameters $\tau_\alpha^0$, $\tau_\beta^0$, $k_\alpha^1$, $k_\beta^1$ were found, the simulation of the first tensile bar was repeated using the finer mesh in figure \ref{fig:meshbcsmall} (c). The model chosen for the hardening of the twin systems was the twin-coplanar model described by equation (\ref{eqn:twinselfhardening}) {as this provides the best optimised fit to the experimental data}. The ``wall'' slip system was the most active because of the lower CRSS, as reported in table \ref{tab:optimizedparameters}. Figure \ref{fig:rhowallbar3} shows the dislocation density $\rho_{1}^\textrm{for}$ in the ``wall'' slip system at 1\% total strain. ``Wall'' slip is mostly active in grain 1 and the dislocation density grows by a factor of approximately 100 compared with the initial value.

\begin{figure}[!htb]
\centering
\includegraphics[width=1.0\textwidth]{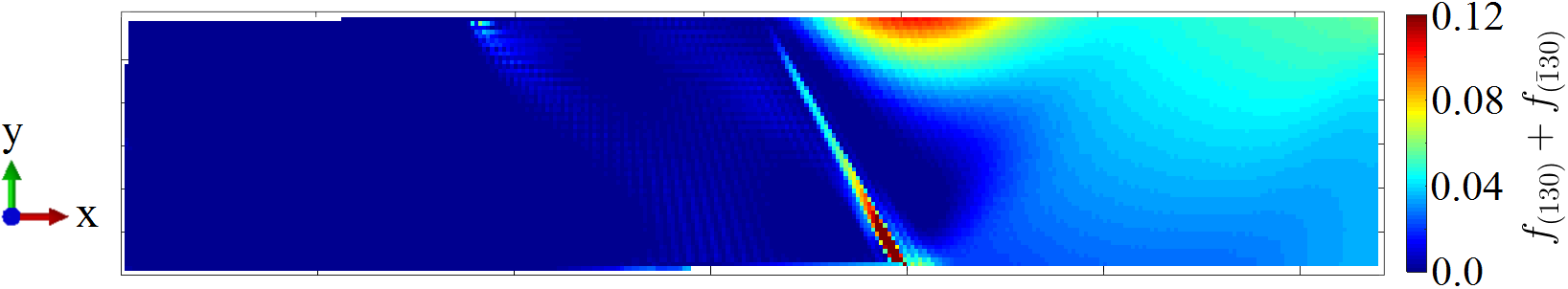}
\caption{\label{fig:twinvolfracbar3}\{130\} Twin volume fraction in the first tensile bar at 2\% total strain {for the twin-coplanar model of equation} (\ref{eqn:twinselfhardening}).}
\end{figure}

Figure \ref{fig:twinvolfracbar3} shows the twin volume fraction $f_\beta$ at 2\% total strain. It grows in the upper part of grain 4 and at the interface between grain 1 and grain 4, reaching a value of about 12\%. We have found that the contribution to the twin volume fraction of the system $\left ( 130 \right ) \left [3\bar{1}0 \right ]$ is much larger than the contribution of $\left ( \bar{1}30 \right ) \left [\bar{3}\bar{1}0 \right ]$. The twin-slip model and the twin-coplanar model lead to the correct prediction of the stress-strain curve in grain 4, as shown in figure \ref{fig:grain1stressstrain} (a) and (c). {The} optimisation procedure allows us to find the interaction coefficients between coplanar twins and between slip and twin systems. 

\begin{figure}[!htb]
\hspace{1.3cm}
\subfloat[(a)]{\includegraphics[width=0.95\textwidth]{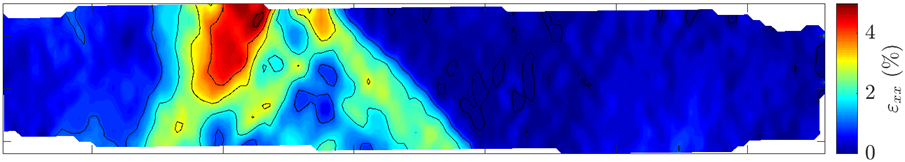}} \\
\subfloat[(b)]{\includegraphics[width=1.05\textwidth]{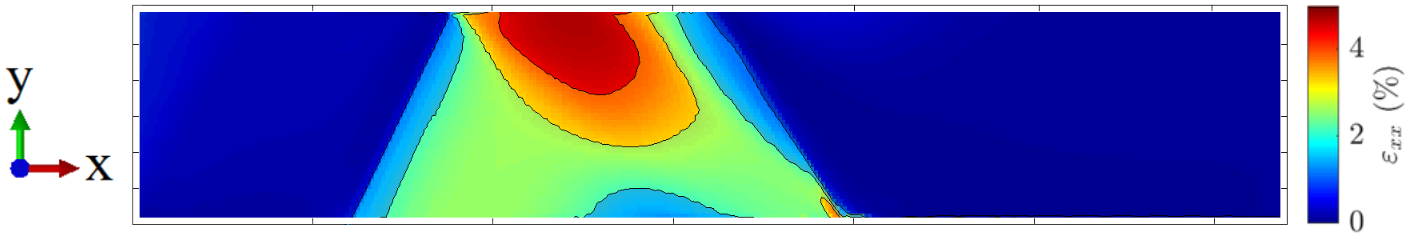}}
\caption{\label{fig:axialstrainTB3}Axial strain $\varepsilon_{xx}$ in the first tensile bar at 1\% total strain: (a) DIC measurement, (b) simulation.}
\end{figure}

\begin{figure}[!htb]
\hspace{1.3cm}
\subfloat[(a)]{\includegraphics[width=0.97\textwidth]{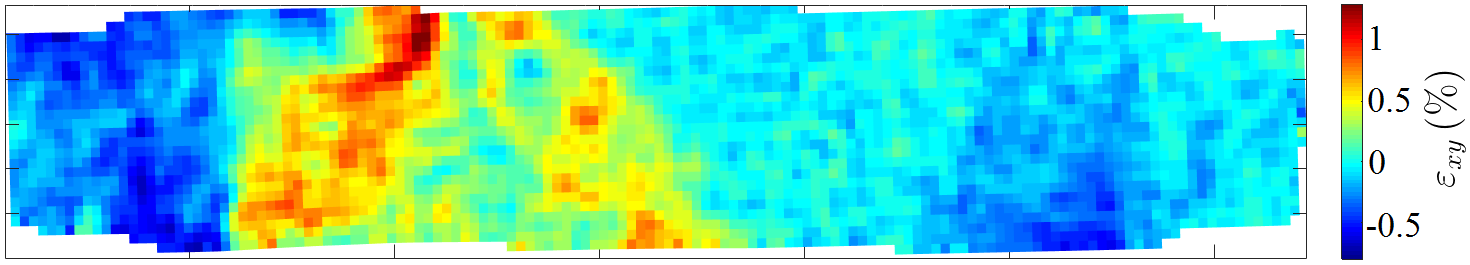}} \\
\subfloat[(b)]{\includegraphics[width=1.05\textwidth]{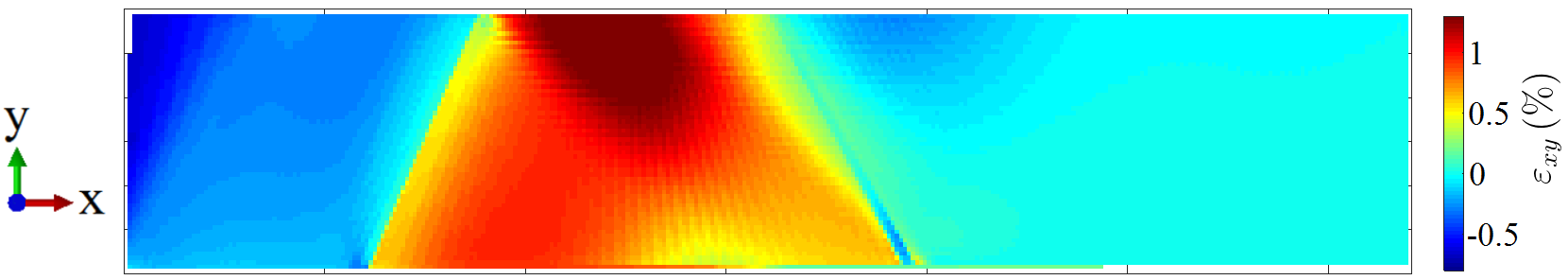}}
\caption{\label{fig:LE12TB3}Shear strain $\varepsilon_{xy}$ in the first tensile bar at 1\% total strain: (a) DIC measurement, (b) simulation.}
\end{figure}

{The detailed strain and displacement fields within the grains provide additional information that can be used to validate the model resulting from the optimisation procedure.} The axial strain concentrates in the upper part of grain 1 due to the activity of the ``wall'' slip system, as shown in figure \ref{fig:axialstrainTB3}. This agrees with the DIC measurement.
The same slip activity leads to the concentration of the shear strain component $\varepsilon_{xy}$ in the upper part of grain 1, as shown in figure \ref{fig:LE12TB3}.

\begin{figure}[!htb]
\subfloat[(a)]{\includegraphics[width=0.46\textwidth]{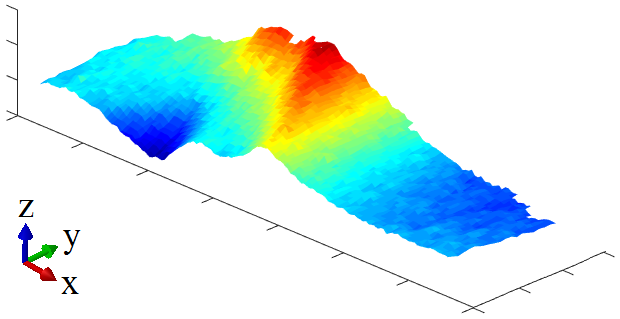}}
\subfloat[(b)]{\includegraphics[width=0.46\textwidth]{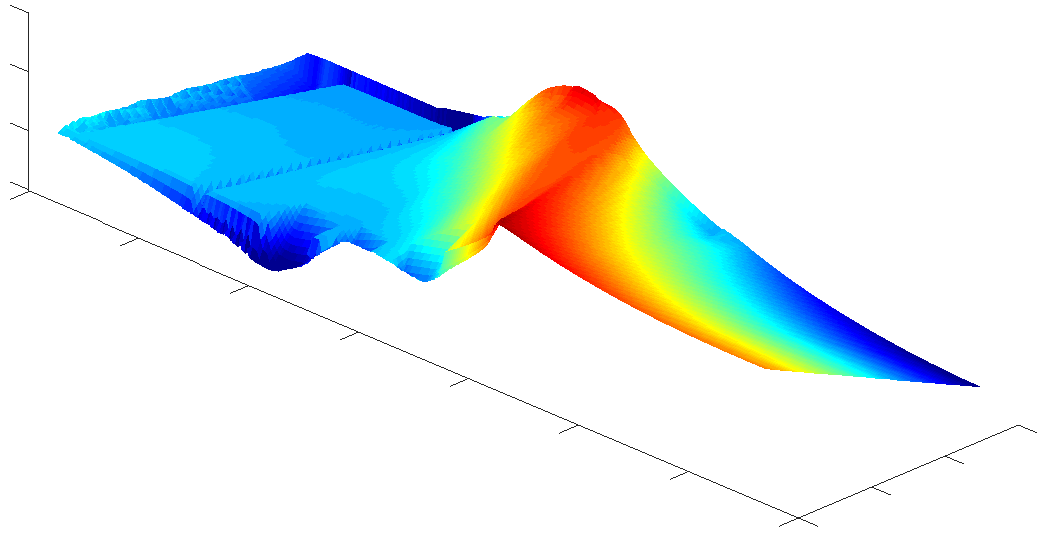}}
\subfloat[]{\includegraphics[width=0.13\textwidth]{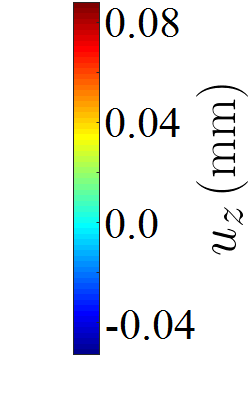}}
\caption{\label{fig:uzbar3}Out-of-plane displacement $u_{z}$ in the first tensile bar at 1\% total strain: (a) DIC measurement, (b) simulation.}
\end{figure}

The measured and predicted out-of-plane displacement $u_z$ is shown as a surface plot in figure \ref{fig:uzbar3}. The simulation results can be interpreted by studying the direction of the Burgers vectors of the ``wall'' slip system and of the $\{130\}$ twin system. The largest component of these vectors is the $u_z$ component, {which is} positive. This is found by multiplying the rotation matrices $\bm{R}$ of grains 1 and 4 and the slip and twin directions $\bm{s}_{\alpha=1}^0$ and $\bm{s}_{\beta=1}^0$ in the lattice reference system. Therefore, the large activity of the ``wall'' slip system in the upper part of grain 1 leads to positive $u_z$ in that region. The same can be stated for the large twin activity on the bottom part of the grain boundary between grain 1 and 4, shown in figure \ref{fig:twinvolfracbar3}. This agrees with the DIC measurement in figure \ref{fig:uzbar3} (a).

{The twin-noncoplanar model described by equations} (\ref{eqn:twintwinhardening})-(\ref{eqn:twintwinhardening2}) {does not predict strong hardening of the most active twin system $\left ( 130 \right ) \left [3\bar{1}0 \right ]$ in grain 4. This explains the underestimation of the maximum stress reached in grain 4 using the twin-noncoplanar model, as shown in figure} \ref{fig:grain1stressstrain} (b).


\begin{figure}[!htb]
\hspace{1.5cm}
\subfloat[(a)]{\includegraphics[width=0.95\textwidth]{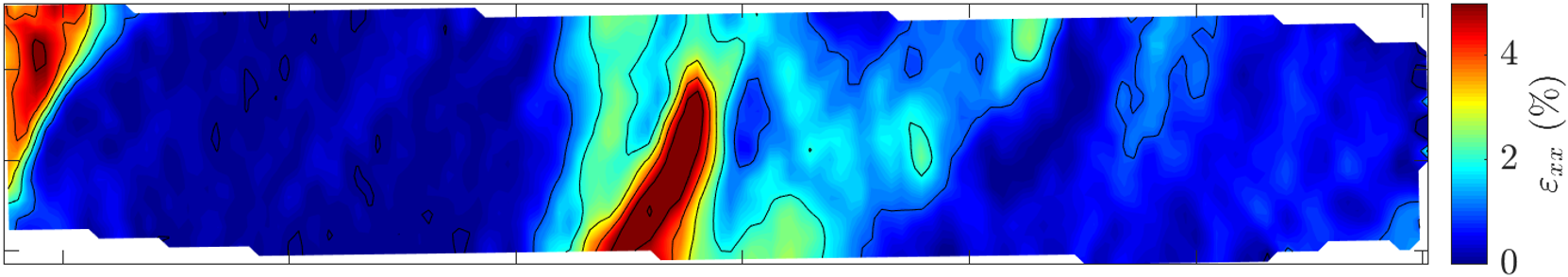}} \\
\subfloat[(b)]{\includegraphics[width=1.04\textwidth]{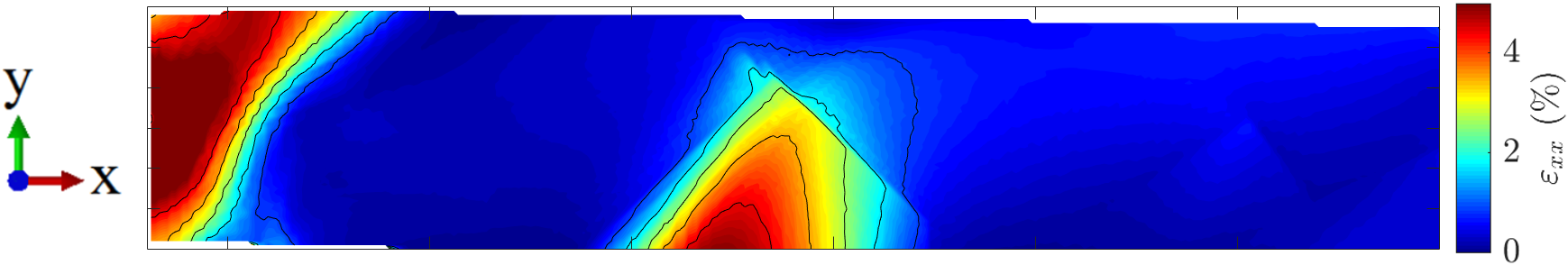}}
\caption{\label{fig:axialstrainTB1}Axial strain $\varepsilon_{xx}$ in the second tensile bar at 1\% total strain: (a) DIC measurement, (b) simulation.}
\end{figure}

\begin{figure}[!htb]
\hspace{1.5cm}
\subfloat[(a)]{\includegraphics[width=0.95\textwidth]{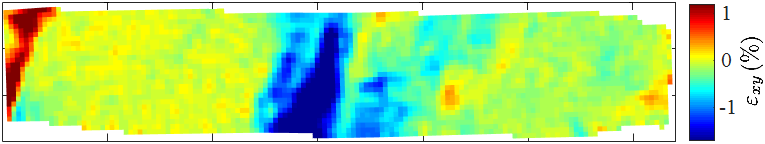}} \\
\subfloat[(b)]{\includegraphics[width=1.04\textwidth]{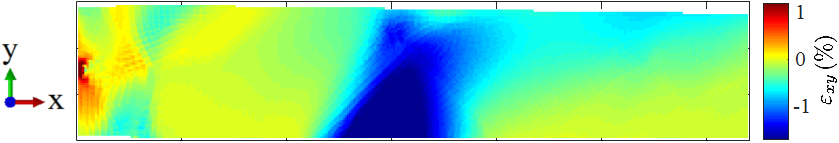}}
\caption{\label{fig:shearstrainTB1}Shear strain $\varepsilon_{xy}$ in the second tensile bar at 1\% total strain: (a) DIC measurement, (b) simulation.}
\end{figure}

An independent model validation was obtained by simulating the second tensile bar with 6 grains shown in figure \ref{fig:TB1EBSD}. A fine mesh with 15984 elements was used. The boundary conditions were the same as for the first tensile bar, shown in figure \ref{fig:meshbcsmall} (a). The same model parameters $\tau_\alpha^0$, $\tau_\beta^0$, $k_\alpha^1$, $k_\beta^1$, optimised for the first tensile bar, were used. The model chosen for the hardening of the twin systems was again the twin-coplanar model described by equation (\ref{eqn:twinselfhardening}). The measured and predicted axial strains are shown in figure \ref{fig:axialstrainTB1}. The ``wall'' slip system is active in grains 1 and 5, leading to strain concentration in those grains.
The same slip activity leads to the concentration of the shear strain component $\varepsilon_{xy}$ in grain 1 and 5, as shown in figure \ref{fig:shearstrainTB1}.

\begin{figure}[!htb]
\subfloat[(a)]{\includegraphics[width=0.46\textwidth]{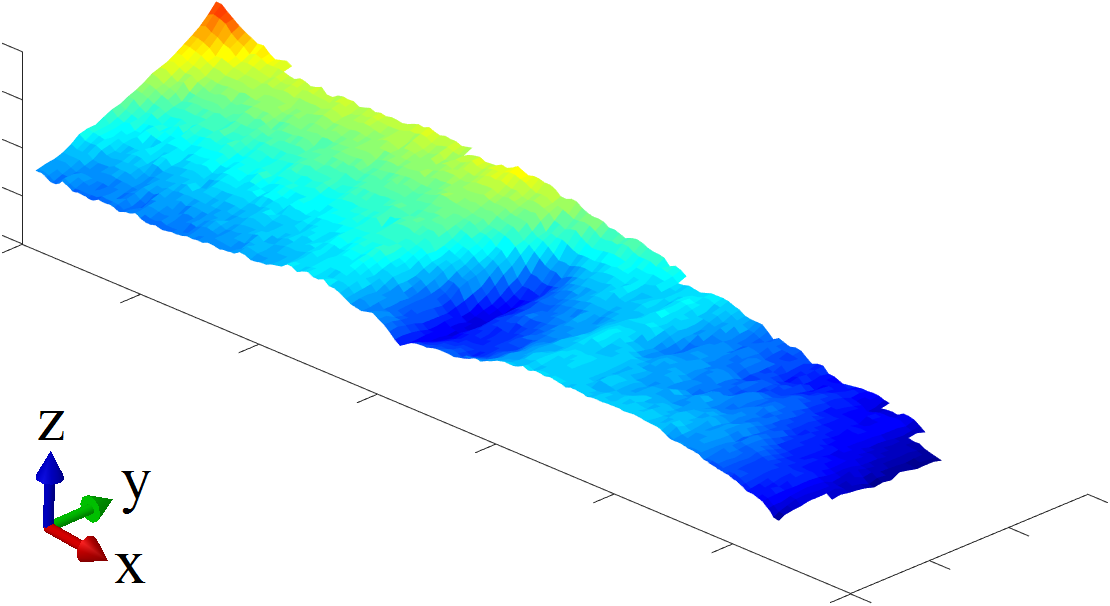}}
\subfloat[(b)]{\includegraphics[width=0.46\textwidth]{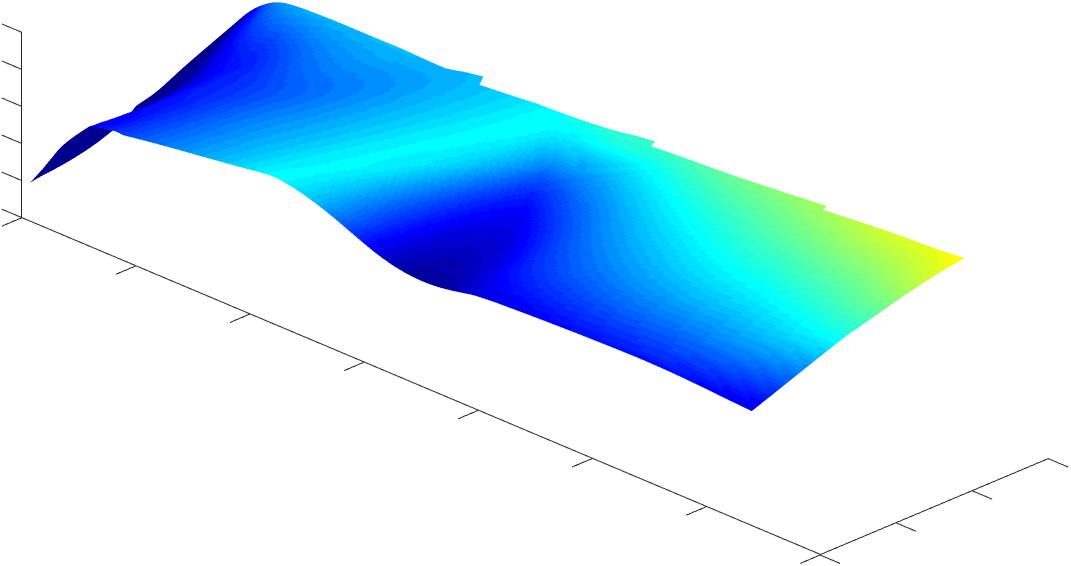}}
\subfloat[]{\includegraphics[width=0.11\textwidth]{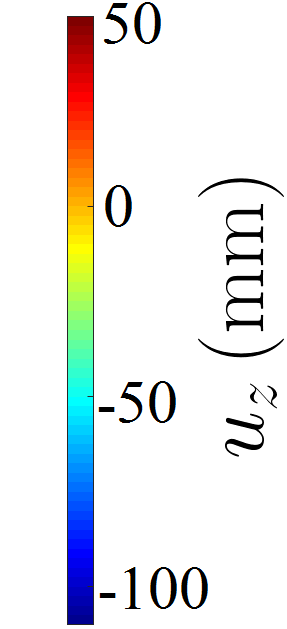}}
\caption{\label{fig:uzbar1}Out-of-plane displacement $u_{z}$ in the second tensile bar at 1\% total strain: (a) DIC measurement, (b) simulation.}
\end{figure}

The measured and predicted out-of-plane displacement $u_z$ is shown in figure \ref{fig:uzbar1}. A rotation of about $1\degree$ around the x axis is added to the simulation results to match the experimental inclination on the surface $x=0$, which was not present in the first tensile bar. Both the measurement and the simulation show a negative region of $u_z$ corresponding to grain 1.

\section{Discussion}
\label{S:discussion}

The optimisation method presented in section \ref{S:optimization} shows that it is possible to use a single tensile bar experiment and DIC measurements to calibrate the yield strength and hardening rate of a crystal plasticity constitutive model for $\alpha$-uranium. This material has slip and twin systems with very different properties. The reliability of the method is confirmed by the correct prediction of the strain and displacement fields, as shown in figures \ref{fig:axialstrainTB3}-\ref{fig:uzbar1}. This optimisation technique can be readily adapted to other materials. 

The stress-strain curves in figure \ref{fig:grain1stressstrain} (a)-(c) show that the \{130\} twin system has a significant strain hardening response. For instance, {the yield strength of} grain 4 in the first tensile bar {increases as the bar is loaded} even though most plastic deformation in that grain {is provided} by a single twin variant $\left ( 130 \right ) \left [3\bar{1}0 \right ]$, as shown in figures \ref{fig:rhowallbar3} and \ref{fig:twinvolfracbar3}. {Early} studies on twin initiation and growth assumed a constant CRSS for twinning \citep{CHRISTIAN19951}. {Recent} experimental {studies} \citep{PATRIARCA2013165} and molecular dynamics simulations \citep{Ojha2014}, {however,} show that both slip activity and twins affect the CRSS for twinning. This is consistent with the present work, in which the optimisation algorithm automatically identifies the {extent of the} twin-slip interaction and the interaction between coplanar twins. The comparison between the twin-slip model and twin-coplanar hardening models, reported in section \ref{S:twinsystemhardening2}, shows that both models can describe the mechanical behaviour of the first tensile bar. It is likely that the twin variant $\left ( 130 \right ) \left [3\bar{1}0 \right ]$ interacts with both slip systems and coplanar twins. The present approach is not able to distinguish between these two interactions, but it can provide values for the interaction coefficients $k_\beta^1$.

The determination of the coefficients $k_\beta^1$ was not possible in previous works on fine-grained $\alpha$-uranium, based on tensile and compression tests of textured samples \citep{McCabeTome2010}, and on neutron diffraction \citep{Brown2009NeutronDiffraction}. This is because these measurements include information from thousands of grains and are not able to discriminate between the slip-slip and slip-twin interactions. Therefore, the interaction coefficients between the slip systems and the $\{130\}$ twin system were set to zero (see table 2 in \citep{McCabeTome2010}). 

Values for the twin hardening coefficients $k_\beta^1$ of $\alpha$-uranium are not available in the literature, but a quantitative comparison with other materials can be made. \citep{AbdolvandDaymond1JMPS2013} {used} a Voce hardening law to describe Zircaloy, in which the CRSS for tensile twins was given in the small strain {approximation:}
\begin{equation}
\tau_\beta^c \approx \tau_\beta^0 + \theta_\beta^0 \Gamma \ ,
\label{eqn:abdolvandvocehard}
\end{equation}
where $\Gamma$ is the accumulated shear on all slip/twin systems and $\theta_\beta^0 = 50$ MPa. Assuming that the $\{130\}$ twin system is the only plastic deformation mechanism, equation (\ref{eqn:twinselfhardening}) for the twin-coplanar hardening model becomes:
\begin{equation}
\tau_\beta^c = \tau_\beta^0 + \left ( \frac{k_\beta^1 \mu_\beta}{\gamma_\beta^\textrm{twin}} \right ) \Gamma \approx \tau_\beta^0 + 323 \Gamma \ ,
\label{eqn:vocetwinselfhardening}
\end{equation}
{where the stress is expressed in MPa.} This shows that the values obtained for $k_\beta^1$ have the same order of magnitude as in other materials. Generally, the hardening rate found in the present experiments is lower than {that} found for fine-grained $\alpha$-uranium. For instance, the dislocation multiplication prefactor of the ``wall'' slip system reported by McCabe et al. \citep{McCabeTome2010} is $k_1^1 / b_1 = 200$ $\mu$m$^{-1}$ while the present optimisation algorithm finds a maximum value of $k_1^1 / b_1 = 43$ $\mu$m$^{-1}$, as reported in tables \ref{tab:modelparameters} and \ref{tab:optimizedparameters}. This is consistent with the finding that the average dislocation density grows faster in fine-grained materials \citep{HAOUALA201872,DeSansal201025}.

Mesoscale models for FeCr single crystals, based on molecular dynamics simulations \citep{Ojha2014}, show that the twin migration stress grows linearly with the residual Burgers vector left at the twin interface during twin-dislocation interaction (see figure 14 in \citep{Ojha2014}). Those simulations are carried out in a single crystal with a characteristic dimension of $500w$, where $w$ is the dislocation core width. A migration stress of 167 MPa is found if the residual Burgers vector is $\left | b_r \right | = a$, where $a =0.286$ nm is the lattice constant \citep{ojhaUIUC}. Assuming $w \approx a$, the residual Burgers vector in those simulations corresponds to a dislocation density $\rho \approx 1 / \left ( 500 a \right )^2 \approx 49$ $\mu$m$^{-2}$. The linear relationship between the dislocation density and the twin migration stress gives an increase of $167$ MPa $/$ $49$ $\mu$m$^{-2}$ $\approx$
 $3.4$ MPa $/$ $\mu$m$^{-2}$. Also equation \eqref{eqn:twinsliphardening} predicts an increase of the CRSS $\tau_{\beta}^{c}$, above which twin volume fraction grows, that depends linearly on the dislocation density; the factor of proportionality is $k_\beta^1 \mu_\beta b_\beta b_\alpha \approx 14$ MPa $/$ $\mu$m$^{-2}$. This comparison shows the consistency between the present model, calibrated without imaging individual twins, and atomistic models, which determine the interaction between dislocations and twins at the sub-micron length scale.

The uncertainty in the present optimisation procedure arises from the approximation that the four grains of the first tensile bar in figure \ref{fig:TB3EBSD} {each} have uniform crystal {orientations.} An uncertainty quantification procedure was carried out: the Euler angles representing the crystal orientation were sampled at different points in the 4 grains; for instance, this leads to an uncertainty of about 20\% on the Schmid factor of the ``wall'' slip system. The optimisation procedure has been repeated with different Euler angles. The uncertainty is small, about 2\%, on the constant friction stress $\tau_1^0$ and on the dislocation multiplication prefactor $k_1^1$ of the ``wall'' slip system, and it is about 30\% {for} $\tau_\beta^0$ and $k_\beta^1$ of the twin system. 

The constant friction stress of the ``chimney'' slip system has a large uncertainty: if the value is changed, the residual $R_\textrm{CRSS}$ of the optimisation procedure does not change significantly. This is because of the low Schmid factor of the ``chimney'' slip system in the 4 grains and the higher CRSS.


\section{Conclusions}
\label{S:conclusions}

An optimisation procedure was developed to find parameters of a dislocation-based constitutive model, implemented in a crystal plasticity finite element framework. {DIC} measurements were {made} during tensile {tests} on coarse-grained $\alpha$-uranium bars containing {a small number of grains}. This allows us to find the average strain in a grain and to plot the stress-strain curves of individual grains, which can be directly compared to the {simulations}.

The constitutive model includes the activity of 8 slip systems and 2 twin variants, as previously identified in $\alpha$-uranium \citep{Zhou2016retwinning}. Three different models for the hardening of the twin systems were compared. {Models which account for} the interaction between twin and slip systems {and} the interaction {between} coplanar twins can explain the experimental data. The optimisation procedure {is} able to find the values of the interaction coefficients, which have not been identified in previous studies on $\alpha$-uranium.

The optimisation procedure {is based on consideration of the average response of the grains within the coarse-grained specimen. The resulting model has been validated by comparing the detailed experimental and simulated strain and displacement fields within the same specimen and for a second tensile specimen with a different grain structure.} These are interpreted by analysing the activity of twin/slip systems and the direction of their Burgers vectors.

The method presented can be applied to find the CRSS and hardening properties of plastic deformation mechanisms of other coarse-grained materials. The method has the advantage that the preparation of single crystal samples is not required {and a large number of parameters can be determined from a single test}.

\section*{Acknowledgements}

The authors acknowledge financial support from AWE plc for this research, program manager: Dr John Askew. ET acknowledges support from the Engineering and Physical Sciences Research Council under Fellowship grant EP/N007239/1. The authors kindly acknowledge Dr Keith Hallam, Christopher Jones and Joseph Sutcliffe for their assistance with the experimental work.





\bibliographystyle{model2-names}
\bibliography{sample.bib}







\end{document}